\documentstyle[11pt,epsfig,aaspp4]{article}
\tighten 
\begin{document}

\newcommand{\lya}{Lyman~$\alpha$}
\newcommand{\lyb}{Lyman~$\beta$}
\newcommand{\za}{$z_{\rm abs}$}
\newcommand{\ze}{$z_{\rm em}$}
\newcommand{\cmtwo}{cm$^{-2}$}
\newcommand{\nhi}{$N$(H$^0$)}
\newcommand{\nzn}{$N$(Zn$^+$)}
\newcommand{\ncr}{$N$(Cr$^+$)}
\newcommand{\degpoint}{\mbox{$^\circ\mskip-7.0mu.\,$}}
\newcommand{\halpha}{\mbox{H$\alpha$}}
\newcommand{\hbeta}{\mbox{H$\beta$}}
\newcommand{\hgamma}{\mbox{H$\gamma$}}
\newcommand{\kms}{\,km~s$^{-1}$}      
\newcommand{\minpoint}{\mbox{$'\mskip-4.7mu.\mskip0.8mu$}}
\newcommand{\mv}{\mbox{$m_{_V}$}}
\newcommand{\Mv}{\mbox{$M_{_V}$}}
\newcommand{\peryr}{\mbox{$\>\rm yr^{-1}$}}
\newcommand{\secpoint}{\mbox{$''\mskip-7.6mu.\,$}}
\newcommand{\sqdeg}{\mbox{${\rm deg}^2$}}
\newcommand{\squig}{\sim\!\!}
\newcommand{\subsun}{\mbox{$_{\twelvesy\odot}$}}
\newcommand{\et}{et al.~}

\def\ltsima{$\; \buildrel < \over \sim \;$}
\def\simlt{\lower.5ex\hbox{\ltsima}}
\def\gtsima{$\; \buildrel > \over \sim \;$}
\def\simgt{\lower.5ex\hbox{\gtsima}}
\def\arcs{$''~$}
\def\arcm{$'~$}
\vspace*{0.1cm}
\title{THE ULTRAVIOLET SPECTRUM OF MS~1512$-$cB58: AN INSIGHT INTO LYMAN 
BREAK GALAXIES\altaffilmark{1}}

\vspace{1cm}
\author{\sc Max Pettini}
\affil{Institute of Astronomy, Madingley Road, Cambridge, CB3 0HA, UK}
\author{\sc Charles C. Steidel\altaffilmark{2} and Kurt L. Adelberger}
\affil{Palomar Observatory, Caltech 105--24, Pasadena, CA 91125}
\author{\sc Mark Dickinson and Mauro Giavalisco}
\affil{Space Telescope Science Institute, 3700 San Martin Drive,
Baltimore, MD 21218}

\altaffiltext{1}{Based on data obtained at the W. M. Keck Observatory which
is operated as a scientific partnership among the California 
Institute of Technology, the
University of California, and NASA, and was made possible by
the generous financial support of the W.M. Keck Foundation. 
}
\altaffiltext{2}{NSF Young Investigator}

\newpage
\begin{abstract}
We present an intermediate resolution, high S/N spectrum of the 
$z = 2.7268$ galaxy MS~1512$-$cB58, obtained with LRIS on the Keck~I 
telescope and covering the rest frame far-UV from 1150 to 1930~\AA.
Gravitational lensing by a foreground cluster boosts the 
flux from cB58 by a factor of $\sim 30$ and provides the opportunity
for a first quantitative study of the physical properties of star forming 
galaxies at high redshift. The spectrum we have recorded 
is very rich in stellar and interstellar features; from their analysis 
we deduce the following main results.

The ultraviolet spectral properties of cB58 are remarkably similar to
those of nearby star forming galaxies and spectral synthesis models 
based on libraries of O and B stars can reproduce accurately
the fine detail of the integrated stellar spectrum. 
The P-Cygni profiles of C~IV and N~V are best matched by continuous star 
formation with a Salpeter IMF extending beyond $M = 50~M_{\odot}$---we 
find no evidence for either a flatter IMF (at the high mass end), 
or an IMF deficient in the most massive stars.
There are clues in our data that the metallicity of both the stars and 
the gas is a few times below solar. 
Our best estimate, $Z_{\rm cB58} \approx  1/4~Z_{\odot}$, 
is $\approx 3$ times higher than the typical metallicity of damped \lya\ 
systems at the same redshift, consistent with the proposal that the 
galaxies which dominate the H~I absorption cross-section are generally
forming stars at a slower rate than $L^*$ Lyman break galaxies like cB58.
The relative velocities of the stellar lines, interstellar absorption, and 
H~II emission indicate the existence of large-scale outflows 
in the interstellar medium of cB58, with a bulk outward motion of 
200~km~s$^{-1}$ and a mass loss rate of $\approx 60~M_{\odot}$~yr$^{-1}$, 
roughly comparable to the star formation rate. 
Such galactic winds seem to be a 
common feature of starburst galaxies at all redshifts, and may well be 
the mechanism which self-regulates star formation, distributes 
metals over large volumes and allows the escape of ionizing photons into 
the IGM.

We suggest further observations of cB58 which should provide more precise 
measurements of element abundances and of detailed physical parameters, and 
highlight the need to identify other examples of gravitationally 
lensed galaxies for a comprehensive study of star formation at early times.
\end{abstract}

\newpage
\section{INTRODUCTION}

Galaxies at $z \simeq 3$ are now being found in large numbers 
through color selection based on the Lyman break and narrow-band 
imaging tuned to the wavelength of \lya\
(e.g. Steidel et al. 1999a,b; Hu, Cowie, \& McMahon 1998). 
The main spectroscopic properties 
of this population were outlined in the original discovery 
paper by Steidel et al. (1996) who showed that Lyman break galaxies
resemble present day star forming galaxies, with spectra 
characterised by a blue ultraviolet continuum with moderate dust 
extinction, strong interstellar absorption, P~Cygni C~IV and N~V lines 
from massive stars, and weak \lya\ emission (see also 
Lowenthal et al. 1997; Meurer et al. 1997; Trager et al. 1997;
Heckman et al. 1998; Pettini et al. 1998b). 

However, most studies of high redshift galaxies have so far concentrated 
mainly on their global properties, such as the luminosity function and
large-scale distribution (e.g. Steidel et al. 1999a; Giavalisco et al. 1998; 
Adelberger et al. 1998). The simple reason is that even with the light 
gathering power of large telescopes the typical galaxy at $z \simeq 3$
is too faint ($L^{\ast}$ corresponds to $R \simeq 24.5$; Steidel et al. 1999a) 
to yield spectra with signal-to-noise ratios of more than a few. 
Thus, more detailed studies of the physical properties 
of these early episodes of star formation require an additional 
observational aid, the light magnification produced by gravitational 
lensing. A few examples have already been identified serendipitously 
and no doubt many more await discovery by targeted searches in
the fields of foreground clusters of galaxies. 

One of the best known cases is the galaxy MS~1512$-$cB58. 
Discovered in the course of the CNOC cluster redshift survey
(Yee et al. 1996), cB58 is exceptionally bright 
(AB$_{6540} = 20.41$, Ellingson et al. 1996) for its redshift
$z = 2.72$\,. The suggestion by William \& Lewis (1997)
that this due to gravitational lensing was confirmed by {\it HST}
images analysed by Seitz et al. (1998) who derived an overall 
magnification of 3.35--4 magnitudes. 
Thus cB58 appears to be a typical $L^{\ast}$ Lyman break galaxy 
fortuitously made accessible to detailed spectroscopic studies by the 
presence of the foreground cluster MS~1512$+$36 at $z = 0.37$\,.

In this paper we present high S/N observations of cB58
(\S2) which offer the best insights yet into the young
stellar population (\S3), interstellar gas and dust (\S4 and \S5), 
and large scale motions (\S6) in a high redshift galaxy.
\S7 deals with intervening absorption.
We summarize our findings in \S8 and highlight the importance 
of extending this type of detailed spectroscopic analysis to
other examples in order to build a comprehensive picture of the 
physical properties of the galaxy population at $z \simeq 3-4$\,.\\

\section{OBSERVATIONS AND DATA REDUCTION}

We used the Low Resolution Imaging Spectrograph (LRIS; Oke et al. 1995)
on the Keck~I telescope on Mauna Kea, Hawaii, 
to record the spectrum of cB58 during two 
observing runs, in 1996 May and August. 
Most of the data were obtained with the 900 grooves~mm$^{-1}$ grating
set to cover the wavelength range 4300--6020~\AA\ with a linear
dispersion of 0.84~\AA~pixel$^{-1}$; the total integration time was 
11\,400~s made up of individual exposures typically 1800~s long.
Additionally, we secured two 1800~s long exposures further to the red, 
using the 1200 grooves~mm$^{-1}$ grating blazed at 7500~\AA\ to
cover the interval 5875--7185~\AA\ at 0.64~\AA~pixel$^{-1}$.
The detector was a SITE $2048 \times 2048$ pixel CCD.
LRIS was used in single-slit mode, with the slit aligned along the
long axis of cB58, which is distorted into a gravitational fold arc 
approximately 3 arcsec long. All the observations were conducted at low 
airmass.

The data were reduced using standard techniques with IRAF.
Since no spatial variations along the arc could be discerned 
(consistent with the gravitationally lensed nature of cB58),
we added up all the signal from the 3 arcsec image in the extraction.
Internal lamps were used for wavelength calibration. 
The spectra were flux calibrated by reference
to spectrophotometric standards and corrected for weak telluric 
absorption by dividing by the spectrum of a B star observed at similar 
airmass. Finally, `blue' and `red' spectra were mapped onto a common
vacuum heliocentric wavelength scale. 
The spectral resolution indicated by the profiles of 
night sky emission lines is
3.0 and 2.1~\AA\ FWHM in the blue and red portions of the 
spectrum, sampled with 3.5 and 2.5 0.85~\AA\ pixels respectively. 
The signal-to-noise ratio per pixel,
measured directly from the final co-added spectrum, is
S/N$\simeq 40$ and $\simeq 15$ in the blue and red respectively.

Figure 1 shows the reduced spectrum. In the rest frame of 
cB58 at $z_{\rm stars} = 2.7268$ (see below) we sample the far-UV 
spectral region, from below \lya\ to just beyond CIII] $\lambda 1909$.
Somewhat ironically, this is one of the best ultraviolet spectra 
of a starburst galaxy obtained at any redshift, including local 
examples studied with {\it HST} such as 
NGC~1741 (Conti, Leitherer, \& Vacca 1996),
NGC~4214 (Leitherer et al. 1996),
and NGC~1705 (Heckman \& Leitherer 1997).
The spectrum is extremely rich in features, mostly originating from 
stars and interstellar gas in cB58, but also from intervening gas from 
the intergalactic medium and galaxies (including the Milky Way) along the 
line of sight. We consider each in turn.\\

\section{THE STELLAR SPECTRUM}
\subsection{Systemic Redshift}

The far-UV emission of a star forming galaxy is 
due primarily to O and B stars. At the 
high S/N of our blue spectrum (top panel in 
Figure 1), most of the low contrast structure seen 
in the continuum is due to stellar features (rather
than noise). In the composite spectrum of a stellar 
population these features are largely blends of different lines
which require stellar population synthesis to be analysed quantitatively
(see below). We have used the spectral atlases by Walborn, 
Nichols-Bohlin, \& Panek (1985) of {\it IUE} observations
of O stars, and by Rogerson \& Upson (1977) and Rogerson \& Ewell (1985)
of the {\it Copernicus} spectrum of the B0~V star $\tau$~Scorpii
to identify stellar photospheric lines which appear to be least affected by 
blending and which can therefore provide a measure of the systemic 
redshift of the stellar population in cB58. These lines are listed in 
Table 1 and are also indicated with tick marks above the spectrum in 
Figure 1. We deduce
$z_{\rm stars} = 2.7268 \pm 0.0008$~($1 \sigma$); the internal 
consistency is reasonable, given the weak, broad character of these features.
S~V~$\lambda 1501.76$ (Howarth 1987) is a prominent line in O stars
but it is possible that there are other contributors to this feature
(S7 in Figure 1), 
since it is apparently detected in the {\it HST} GHRS spectrum of
the `postburst' NGC~1705-1, where a significant population of O stars 
is no longer present (Heckman \& Leitherer 1997). Its wavelength in cB58 
may be slightly discordant from those of the other photospheric lines in 
Table 1, although the evidence is not clear-cut.

\subsection{Spectral Synthesis}

In the last few years modelling of the integrated
spectra of star forming regions has progressed significantly 
and has been shown to be a powerful tool for deducing many properties of 
the underlying stellar populations, including the age of the burst, 
the Initial Mass Function (IMF), metallicity, and dust reddening
(e.g. Leitherer 1996). In applying the technique to cB58 we made use 
of {\it Starburst99\/}, the comprehensive set of models
recently compiled by Leitherer et al. (1999).

It is evident when considering these models that `continuous 
star formation' provides a much better description of the spectrum
of cB58 than `single burst' models. We do not see in our data any
features which are indicative of a particular phase in the evolution 
of a starburst, such as WR features (e.g. He~II ~$\lambda 1640$)
and strong Si~IV~$\lambda 1397$ P~Cygni profiles 
which are most prominent after a few Myr,
nor the {\it lack} of C~IV~$\lambda 1549$ and N~V~$\lambda 1240$
P~Cygni lines which signals the absence of 
O stars $\sim 10$~Myr after the burst.
Consequently, we restrict ourselves to continuous star formation 
models in the analysis that follows.
A corollary of this conclusion is that 
it is not possible to derive an age for the galaxy from the rest-frame 
ultraviolet spectrum alone, which is always
dominated by the light from the 
youngest members of the stellar population.
Ellingson et al. (1996) used the broad spectral energy distribution 
from optical and infrared photometry (1300--6000~\AA\ in the rest frame)
to deduce an age of less than 35 Myr, but this conclusion is dependent on 
the amount of reddening which is somewhat uncertain, 
as discussed in \S5 below.

The wavelength regions of most interest for spectral synthesis
are those encompassing the C~IV~$\lambda 1549$ and 
N~V~$\lambda 1240$ lines; in Figures 2 and 3 respectively we compare 
them with the predictions of different {\it Starburst99\/} models,
all for a 20~Myr old continuous star formation 
episode.\footnote{The model spectra shown are the {\it rectified}
versions, in which the stellar continuum has been divided out.
In these continuous star formation models there is only 
a minor dependence of the stellar features on the 
age of the star formation episode.}
Before discussing these figures it is important to clarify 
that the comparison between models and observations involves 
only the stellar lines and that no attempt is made to fit the 
{\it interstellar} lines in the spectrum. 
The {\it Starburst99\/} models are constructed
from libraries of {\it IUE} spectra of Galactic (and therefore relatively 
nearby) O and B stars; in general these stars have much weaker 
interstellar absorption than that seen through an entire galaxy.
In Figures 2 and 3 most of the interstellar lines can be  
recognized from their narrower widths or by reference to Table 2 below.
With this point in mind, it is evident from Figure 2 that the models
provide a remarkably good fit to the broad spectral features in 
the C~IV region. Note that {\it no} adjustment was made in order to match 
observations and models, other than dividing the observed wavelength scale
by the value of ($1 + z_{\rm stars}$) deduced above. 

Focussing on the C~IV line itself, we note that it consists of three 
components. Two components make up the P~Cygni stellar line: 
redshifted emission and a broad 
absorption trough which extends to 
1534.25~\AA\  corresponding to a terminal wind velocity 
$v_{\infty} = -2800$~km~s$^{-1}$
after correcting for the instrumental resolution. 
Superimposed on the stellar line are the narrower 
C~IV~$\lambda\lambda 1548.195, 1550.770$
interstellar doublet lines which reach nearly zero residual intensity.
The profile of the stellar C~IV line can be used to place limits on the 
IMF in this distant star forming galaxy, as shown in Figure 2.
In the top panel we see that a Salpeter IMF, with slope
$\alpha = 2.35$ and upper mass limit $M_{\rm up} = 100~M_{\odot}$
reproduces well the emission component of the P~Cygni profile
(although it overpredicts the optical depth of the absorption 
trough---but see below). 
If $M_{\rm up}$ is reduced to $30~M_{\odot}$
the P~Cygni emission is lost altogether (middle panel); indeed the 
existence of a P~Cygni C~IV profile in itself implies 
that O stars with masses greater than $50~M_{\odot}$
must be present (Leitherer, Robert, \& Heckman 1995).
A similar discrepancy between the observed and predicted 
C~IV profiles is found if a steeper IMF ($\alpha  = 3.3$) is adopted 
(while maintaining $M_{\rm up} = 100~M_{\odot}$).
Conversely, a flatter IMF ($\alpha  = 1.0$),
as proposed for example by Massey, Johnson, \& DeGioia-Eastwood (1995),
greatly overproduces C~IV emission (bottom panel of Figure 2).
As can be seen from Figure 3, similar considerations apply to the N~V 
line (which is also a blend of stellar P~Cygni emission-absorption, 
and interstellar $\lambda\lambda 1238.821, 1242.804$ absorption).
We conclude that there is no evidence of a departure from a Salpeter IMF
(at least at the high mass end) in the star formation episode
taking place in cB58.

We now turn to the strength of the P~Cygni absorption which, 
as noted above, is observed to be weaker in cB58 than in the best fitting 
{\it Starburst99\/} model.   
The optical depth of the trough is sensitive to
the mass loss rate (e.g. Lamers, et al. 1999); for a star of a given 
spectral and luminosity class the mass loss rate decreases
with decreasing metallicity (e.g. Puls et al. 1996). 
Comparisons of O and B stars in the Milky Way and in the  
Magellanic Clouds (e.g. Walborn et al. 1995; Lennon 1999) 
have shown a clear trend of decreasing
strength of C~IV absorption as the carbon abundance
decreases from $\sim 2/3$, to $\sim 1/4$, and to
$\sim 1/7$ of the solar value,  from the Milky Way near the Sun, to the 
LMC, and to the SMC respectively. The trend is most obvious in main 
sequence stars and an 
analogous effect is seen in Si~IV and N~V.
Theoretically, one may expect a metallicity dependence of the 
mass loss rate of the form 
$\dot{M} \propto (Z/Z_{\odot})^{\beta}$, with
$\beta$ in the range $0.5 - 1$ (see eq. 8.63 of Kudritzki 1998).

It thus seems at least plausible that the difference between observations 
and model in the top panels of Figures 2 and 3 is an indication
that the metallicity of the OB stars in cB58 is lower than 
that of the solar neighbourhood stars 
which make up the libraries of stellar spectra of {\it Starburst99\/}.
In the future, when the {\it Starburst99\/} data base is extended to include 
stars in the Magellanic Clouds, 
it may be possible to calibrate empirically the metallicity dependence
of this effect and use spectral synthesis techniques
to measure the metal abundance of high redshift galaxies. 
For the moment, our best 
estimate of the metallicity of cB58 is that it is below solar, and 
comparable to that of 
the Magellanic Clouds.
The preliminary results by Robert et al. (1999,
in preparation), as described by Leitherer (1999),
would favour a value closer to that of the LMC on the basis
of the Si~IV P-Cygni profile which is nearly absent in SMC stars,
but is still present in cB58 (see Figure 1).

In summary, the stellar spectrum of cB58, when compared with the  
predictions of the best available spectral synthesis models, leads us to 
conclude that: (a) this galaxy is undergoing a protracted period of star 
formation; (b) there is no evidence for departure from a Salpeter IMF 
extending to $M_{\rm up} = 100~M_{\odot}$; and (c) the metallicity is 
below solar by a factor of a few.

Finally, we draw attention to the fact that the 
Al~III~$\lambda\lambda 1854,1862$ doublet lines may also include
a stellar wind component, as they seem to exhibit asymmetric blue
wings not dissimilar from that seen in C~IV~$\lambda 1549$ (see
Figure 1). The Al~III doublet can be very strong in early B-type
supergiants, so that a contribution to the integrated spectrum of
cB58 is not implausible.
With better data and more
extensive stellar libraries it should be possible in future to
use this spectral feature, together with C~IV, Si~IV and N~V, to
refine further the spectral synthesis modelling of young stellar
populations.\\

\section{INTERSTELLAR ABSORPTION LINES}

As can be seen from Figure 1, 
the spectrum of cB58 is dominated by interstellar lines.
We identify 29 lines, listed in Table 2, due to elements ranging
from hydrogen to nickel in ionization stages ranging from C~I to N~V.
The internal redshift agreement is excellent; 
we deduce a mean absorption redshift
$z_{\rm abs} = 2.7242 \pm 0.0005$ ($1 \sigma$).
The lines are very strong, 
indicating that the absorption takes place over a 
wide velocity interval; for example the FWHM of Si~II~$\lambda 1526$
and Al~II~$\lambda 1670$ imply a velocity spread of 530~km~s$^{-1}$
(after correction for the instrumental resolution).
Evidently, the interstellar medium of this galaxy has been stirred
to high speeds presumably by the mechanical energy deposited by the 
massive stars through stellar winds and supernovae.

As discussed below (\S6), the \lya\ line in cB58 includes a
damped absorption component with 
$N$(H~I)$= 7.5 \times 10^{20}$~cm$^{-2}$. 
At first glance the interstellar absorption spectrum of cB58 is
not dissimilar from that of many damped \lya\ systems (DLAs),
although there are notable differences. The compilation of
line profiles by Prochaska \& Wolfe (1999) provides a useful
comparison. The widths of the low ionization lines are greater 
in cB58 than in most DLAs, where Si~II~$\lambda 1526$
and Al~II~$\lambda 1670$ are seldom wider than 200~km~s$^{-1}$.
Most significantly, absorption lines from the fine structure
levels of the ground state of Si~II, which are rarely detected in
DLAs, are exceedingly strong
in cB58, with rest frame equivalent widths $W_0 \simeq 0.5$~\AA\
(lines 10 and 20 in Table 2). These levels are 
populated by collisions with electrons and hydrogen atoms
(Keenan et al. 1985), so that the absorption lines we see
must be formed in gas of higher density
than that sampled by DLAs in random sight-lines
to background QSOs.
In our Galaxy unusually prominent fine structure lines have been
seen in regions of violent star formation, such as the Carina nebula
(Laurent, Paul, \& Pettini 1982), 
and in interstellar clouds compressed by the
passage of a supernova induced shock, as in the Vela supernova
remnant (Jenkins \& Wallerstein 1995); finding such strong
Si~II$^*$ absorption in cB58 is therefore not surprising.
With higher resolution observations it may be possible (depending
on line saturation) to measure the 
ratio $N$(Si~II$^*$)/$N$(Si~II) and deduce the electron density
$n_e$.

The resolving power of our spectrum is about one order of magnitude 
lower than that normally required to measure element abundances from 
interstellar absorption lines. 
Nevertheless, it is instructive to consider 
the values which result if we make some simple assumptions.
We restrict ourselves to the weakest lines in the spectrum, 
with rest frame equivalent width $W_0 \leq 0.5$~\AA, and assume that 
saturation effects are unimportant to derive the ion column densities 
$N$ listed in column (6) of Table 3. For Ni~II we use the latest 
$f$-values from the radiative lifetime measurements
by Fedchack \& Lawler (1999), and scale the earlier determinations
by Morton (1991) and Zsarg\'{o} \& Federman (1998) accordingly
(a reduction by a factor of 1.9).
The five Ni~II transitions covered give
internally consistent values of $N$(Ni~II).
With $N$(H~I)$= 7.5 \times 10^{20}$~cm$^{-2}$
deduced in \S6 below and
the assumption that most of the Si~II, S~II and Ni~II
are associated with the H~I gas, we deduce the abundances 
in column (7) of Table 3, and from these arrive 
at the abundances relative to 
solar given in column (9), in the usual notation.

The abundances derived are between 1/3 and 1/5 of solar, 
ostensibly in good 
agreement with our earlier conclusion 
from the analysis of the stellar spectrum.
(\S3). However, until higher resolution observations are 
available, these interstellar estimates remain highly uncertain. 
On the one hand, we 
may have overestimated the metallicity if the H~II gas 
along the line of sight accounts for a significant proportion 
of the first ions (see \S5 below).
On the other hand, if any saturated components contribute to the 
absorption line equivalent widths, the abundances in Table 3 are 
underestimates. Some of these corrections probably apply because
we would have expected some dust depletion of Ni 
(a refractory element) relative 
to S (Savage \& Sembach 1996), whereas none is seen. \\

\section{DUST EXTINCTION}

The spectra of O and early B stars continue to rise in the far-UV
peaking near 1000~\AA\ (e.g. Hubeny \& Lanz 1996).
Model predictions of the integrated light from galaxies 
with on-going star formation
show that in the region between 1800 and 1250~\AA\
the continuum can be approximated by a power law of the form
$F_{\nu} \propto \nu^{\alpha}$, with $\alpha \simeq 0.5$
for a range of metallicities and ages. The {\it Starburst99\/} model 
used here (continuous star formation, 20~Myr, Salpeter IMF, solar 
metallicity) predicts $\alpha = 0.4$\,.
In contrast, it can be seen from Figure 1 that the continuum in 
cB58 (in $f_{\nu}$ units) decreases with decreasing wavelength;
we measure $\alpha = -1.2$\,.\footnote{This slope is consistent
with the broad band photometry of Ellingson et al. (1996).}
Given the presence in the spectrum of discrete features from OB stars 
and of strong interstellar metal lines, the most straightforward 
interpretation of this difference between observations and model 
predictions is that the UV continuum is reddened by dust 
extinction.

To make further progress it is necessary to make
some assumptions about the 
unknown properties of dust in cB58.
One possibility is to use the Magellanic Clouds as 
guidelines, given the similarity in metallicity 
deduced above. Adopting the 
wavelength dependence of the ultraviolet extinction 
of the LMC (Fitzpatrick 1986) with the normalization
by Pei (1992), we find that 
\begin{equation}
      \Delta \alpha = 6.8 \times E{\rm (}B-V{\rm )}
\end{equation}
and 
\begin{equation}
      A_{\rm 1500} = 8.8 \times E{\rm (}B-V{\rm )}
\end{equation}
where $\Delta \alpha$ is the difference between observed and predicted 
spectral slopes in the interval 1250--1800~\AA\
and $A_{\rm 1500}$ is the dust extinction in magnitudes 
at 1500~\AA. With these parameters we deduce 
$E{\rm (}B-V{\rm )} = 0.24$ and $A_{\rm 1500} = 2.1$~mag (a factor of 
$\sim 7$).
The steeper UV rise of the SMC reddening curve (Bouchet et al. 1985)
leads to
\begin{equation}
      \Delta \alpha = 15.9 \times E{\rm (}B-V{\rm )}
\end{equation}
and 
\begin{equation}
      A_{\rm 1500} = 12.6 \times E{\rm (}B-V{\rm )}
\end{equation}
and therefore to the lower estimates, 
$E{\rm (}B-V{\rm )} = 0.10$ and $A_{\rm 1500} = 1.3$~mag.
The extinction properties of dust associated with
the 30~Dor giant H~II region in the 
LMC (Fitzpatrick 1986) would yield values intermediate between these two
possibilities, while the attenuation law derived by Calzetti (1997) for 
local starburst galaxies would give
$E{\rm (}B-V{\rm )} = 0.29$ and $A_{\rm 1500} = 3.3$~mag.
This value of $E{\rm (}B-V{\rm )}$ is in the upper quartile
of the distribution for the whole sample 
considered by Steidel et al. (1999a); evidently cB58 is 
among the more reddened Lyman break galaxies.

Locally, there is a close correlation between the column 
densities of gas and dust which in the Milky Way takes the form 
\begin{equation}
      \langle N{\rm (H~I)}/E{\rm (}B-V{\rm )} \rangle
      = 4.93 \times 10^{21}~{\rm cm^{-2}~mag^{-1}}
\end{equation}
with a standard deviation of approximately 0.19~dex in the sample
of 392 OB stars compiled by Diplas \& Savage (1994).
Although based on a much smaller number of measurements,
it seems well established that the gas-to-dust ratios are larger
than this value in the LMC and the SMC, by factors of 
$\sim 2-4$ and $\sim 10$ respectively (Fitzpatrick 1989), 
presumably reflecting the lower metallicity of these galaxies.
Thus, the values of $E{\rm (}B-V{\rm )}$ deduced above
would imply neutral hydrogen column densities 
$N$(H~I)$\simeq 2.5 \times 10^{21}$ and $5 \times 10^{21}$~cm$^{-2}$
for LMC and SMC conditions respectively.

These values are $\sim 3-7$ times larger than that measured from the 
damped profile of the \lya\ absorption line (see \S6 below); expressed in 
a different way, we apparently see a larger dust-to-gas ratio than 
expected, by a factor of several.
Perhaps it is unrealistic to expect better internal agreement
in such estimates
given the many assumptions involved, and it is certainly possible that we 
have overestimated the dust exctinction 
(if the {\it Starburst99\/} models predict a continuum slope that is too 
blue), or understimated the metallicity. On the other hand, this 
discrepancy may be telling us that more than 2/3 of the gas in  
front of the stars is not in atomic form, but rather is ionized and/or
molecular hydrogen.

The values of dust extinction derived above help us estimate the 
star formation rate of cB58 from its UV continuum luminosity.
Ellingson et al. (1996) reported AB$_{5500} = 20.64 \pm 0.12$ which
at $z_{\rm stars} = 2.7268$ implies 
$L_{1476} =  3.44 \times 10^{30}$~erg~s$^{-1}$~Hz$^{-1}$ 
($H_0 = 70$~km~s$^{-1}$~Mpc$^{-1}$; $q_0 = 0.1$).
The Bruzual \& Charlot (1996, private communication) models provide a 
calibration of $F_{\nu}$ in terms of the star formation rate; assuming
an IMF with the Salpeter slope down to $0.1~M_{\odot}$, 
solar metallicity, and continuous star formation over a 
period of 100~Myr, ${\rm SFR} = 1~M_{\odot}$~yr$^{-1}$
produces $L_{1500} = 8.7 \times 10^{27}$~erg~s$^{-1}$~Hz$^{-1}$
(the dependence of this scaling 
on metallicity and age is a small effect compared with
the other uncertainties discussed below).
From this we deduce a `best value' 
\begin{equation}
{\rm SFR}_{\rm cB58} = 37 \times 
\left( \frac{30}{f_{\rm lens}} \right)  \times 
\left( \frac{f_{\rm dust}}{7} \right)  \times
\left( \frac{2.5}{f_{\rm IMF}} \right)~
M_{\odot}~{\rm yr}^{-1}
\end{equation}
where the values in brackets are correction factors 
respectively for gravitational lens amplification (Seitz et al. 1998), 
dust extinction (this work), and 
the IMF (Leitherer (1998) has proposed 
that estimates of SFR based on 
an extrapolation of the Salpeter IMF down to $0.1~M_{\odot}$
should be reduced by a factor of 2.5 to account 
for the observed flattening of the 
IMF below $1~M_{\odot}$, e.g. Sirianni et al. 1999; Zoccali et al. 1999). 

The `raw' value of SFR implied by the UV continuum luminosity,
without any of the above corrections, 
SFR$^{\prime} = 395~M_{\odot}$~yr$^{-1}$,
is $\sim 3$ times higher than 
SFR$ = 120 \pm 20~M_{\odot}$~yr$^{-1}$
(for the cosmology used here)
deduced by Bechtold et al. (1997) from narrow-band imaging
in H$\alpha$, which is redshifted
near the edge of the infrared $K$-band window at
2.4426~$\mu$m. 
These authors considered several possible
explanations for the discrepancy, including 
higher obscuration of the
emission line gas, 
leakage of ionizing photons from the H~II region,
time-dependent ionization effects, and
differences in the degree of gravitational magnification
across the source. An additional,
simpler, possibility is that the narrow-band 
observations
may have underestimated the true H$\alpha$ luminosity.
An $H$-band spectrum recorded by G. Wright (1999, private
communication) with CGS4 on UKIRT shows a clear continuum and an 
H$\beta$ emission line with an integrated flux
of $1.0 \times 10^{-15}$~ergs~s$^{-1}$~cm$^{-2}$.
If we assume a ratio H$\alpha$/H$\beta = 2.75$ (Osterbrock 1989),
the predicted H$\alpha$ flux is $\sim 5$ times higher
than the value 
($5.8 \pm 1$)$\times 10^{-16}$~ergs~s$^{-1}$~cm$^{-2}$ reported by
Bechtold et al. (1997); 
dust extinction in the Balmer lines
would increase the difference further.
The H$\beta$ flux measured with UKIRT
is in good agreement with the luminosity and reddening
of the far-UV continuum 
derived above, as found for other
Lyman break galaxies (Pettini et al. 1998a).\\

\section{Ly$\alpha$ AND LARGE SCALE OUTFLOWS}

The \lya\ line in cB58 is a blend of absorption and emission.
In Figure 4 we show our decomposition of this feature. 
The damping wings are well fitted with a column density
$N$(H~I)$ = 7.5 \times 10^{20}$~cm$^{-2}$ centred at
$z_{\rm abs} = 2.7240$, in good agreement with the redshifts of the other 
interstellar absorption lines (see Table 2).
Subtraction of the damped \lya\ absorption then reveals a redshifted 
\lya\ emission line (bottom right-hand panel of Figure 4), exhibiting
a highly asymmetric shape with a peak near $+450$~km~s$^{-1}$,
a sharp drop on the blue side, and a tail of emission 
which apparently extends to beyond 1000~km~s$^{-1}$
(velocities relative to $z_{\rm stars}$).
This profile is remarkably similar to that seen in another bright Lyman 
break galaxy, Q0000$-$263~D6 (see Fig.8 of Pettini et al. 1998b).

More generally, redshifted \lya\ emission is often seen in
high redshift galaxies (Pettini et al. 1998a and references therein),
and in local H~II and starburst galaxies (Kunth et al. 1998; Gonz\'{a}lez 
Delgado et al. 1998).
The explanation commonly put forward involves
large scale outflows in the interstellar 
media of the galaxies observed. 
In this picture \lya\ emission is suppressed by 
resonant scattering and the only \lya\ photons which can escape 
unabsorbed in our direction are those 
back-scattered from the far side of the expanding 
nebula, whereas in absorption against the stellar continuum
we see the approaching part of the outflow.

The data presented here are consistent with this scenario and 
indeed provide a better measurement of the velocity fields involved than 
previous observations. Before discussing the kinematics further, we 
draw attention to a number of weak emission lines which can be recognized
from close inspection of Figure 1 and which are listed in Table 4. 
While weak, these features are undoubtedly real
(they are significant at the many $\sigma$ level---see column (6) of 
Table 4); we tentatively interpret them as 
recombination lines to the fine structure 
levels of the ground states of C~II and Si~II, presumably arising in 
an H~II region. The internal redshift agreement is acceptable, as can be 
seen from column (5) of Table 4. The
corresponding resonance lines, which are not detected, are
intrinsically weaker and are subject to strong
absorption by foreground gas, as is \lya.
Such radiation transfer effects in C~II~$\lambda 1335$ have been seen
in low-excitation planetary nebulae (e.g. Clavel, Flower, \&
Seaton 1981).

Interestingly, we also see N~IV]$\lambda 1486.496$. This is normally a 
stellar line, and indeed it can be recognized in the {\it Starburst99\/} 
models reproduced in Fig. 2. However, in cB58 its measured wavelength
seems to agree better (although the line is noisy) 
with the redshift of the other weak emission lines in Table 4 
than with that of the stellar photospheric lines in Table 1.
The spectrum in Figure 1 also shows evidence for C~III]~$\lambda 1909$
emission, but the existing data are too noisy for a reliable measurement 
of this feature. 

Table 5 summarizes the velocity measurements.
Evidently, the interstellar medium in cB58 is expanding with
a bulk velocity of $\sim 200$~km~s$^{-1}$; the \lya\ line, which is most 
sensitive to resonant scattering effects, picks out the gas at the 
highest velocities. 
Our measurements support the broad picture of the kinematics 
of starburst galaxies considered most recently by Tenorio-Tagle et al. 
(1999) in which   
the mechanical energy deposited by the 
massive stars and supernovae 
leads to the formation of a cavity filled with hot ejecta and
surrounded by an expanding shell of swept up interstellar material.
The ensuing 
outflows, which seem to be a common feature of 
starburst galaxies, have potentially several important consequences. 
First, they provide the feedback 
required for self-regulation of the 
star-formation activity. Second, they are a mechanism  
for distributing the 
products of stellar nucleosynthesis over large volumes. Third, 
they may lead to 
the escape of Lyman continuum photons from the galaxies, if the 
cavity created by the expanding superbubble breaks through the ISM, 
with important consequences for the 
ionization of the intergalactic medium at 
high redshift (Madau, Haardt, \& Rees 1999).

We can estimate the mass loss rate involved by considering the flow of 
mass through a unit area (assuming spherical symmetry):
\begin{equation}
\dot{M} = 4 \pi r^2\, n\, v\, m_p
\end{equation}
where $r$ is the radius of the superbubble, 
$n$ is the matter density,
$v$ is the speed of the outflow
and $m_p$  is the mean particle mass.
If we assume that all the material within the superbubble
has been swept-up into a shell of thickness $\Delta r_s$
and density $n_s$, we can substitute the column density
\begin{equation}
N = n_s  \times \Delta r_s = \frac{n \times r}{3}
\end{equation}
in eq. (7) to obtain
\begin{equation}
\dot{M} = 12 \pi\, r\, N\, v\, m_p
\end{equation}
where all the quantities on the right-hand side are measured, except for 
the radius of the superbubble. Adopting $r = 1$~kpc as a working value
(Tenorio-Tagle et al. 1999), we obtain
\begin{equation}
\dot{M}  \simeq 60 \times 
\left( \frac{r}{1~{\rm kpc}} \right)  \times 
\left( \frac{N}{7.5 \times 10^{20}~{\rm cm}^{-2}} \right)  \times
\left( \frac{v}{200~{\rm km~s}^{-1}} \right)~
M_{\odot}~{\rm yr}^{-1}
\end{equation}
or three times higher if neutral hydrogen accounts for only 1/3 of the total 
column of gas in front of the stars, as discussed above (\S5).

Thus we find that, within the uncertainties,
the mass loss rate due to the galactic 
superwind is comparable to
the rate at which gas is 
turned into stars (eq. (6)).
It remains to be established what the fate of the outflowing 
interstellar material is, that is whether it leaves the galaxy 
altogether, or remains trapped in its potential well until it can cool 
and rain back onto the galaxy.
If the initial estimates of the masses of Lyman break galaxies by 
Pettini et al. (1998a) are typical ($M \simgt 10^{10}~M_{\odot}$), 
the gas may well remain bound (Ferrara \& Tolstoy 1999).\\

\section{INTERVENING ABSORPTION LINES}
The spectrum of cB58 shows a number of narrow absorption 
lines produced by gas along the line of sight to this high redshift 
galaxy. A rich \lya\ forest is evident in Figure 1 shortwards of 
1200~\AA; we have not attempted to measure wavelengths and equivalent 
widths of individual absorption features in the forest 
because at the resolution of 
the present data they are all blends.
Intervening lines longwards of \lya\ are listed in Table 6 and marked 
in Figure 1.
We identify two Mg~II absorption systems at $z_{\rm abs} = 0.8287$ and 
1.3391 respectively. The latter is likely to have
$N$(H~I)$ \simgt 1 \times 10^{20}$~cm$^{-2}$ on the basis
of the relatively large value of the 
$W$(Fe~II~$\lambda 2382$)/$W$(Mg~II~$\lambda 2796$) ratio
(Bergeron \& Stasi\'{n}ska 1986); note also the strength of
Mg~I~$\lambda 2852$. 
It will be interesting to search for galaxies at 
these redshifts in deep images of the cluster MS~1512+36.
Our spectrum does not cover any strong absorption lines 
which may be due to the cluster itself, at $z = 0.373$ (Gioia \& Luppino 
1994); at this redshift the Mg~II doublet would fall at 3841.5~\AA.
We see no absorption at the wavelength of Ca~II~$\lambda 3934.777$
to a formal equivalent width limit $W_0$($3 \sigma$)$ = 0.12$~\AA. 

We apparently detect Na~I~$\lambda 5891$ (the Na~D2 line) from the disk 
and halo of our own Galaxy, although its anomalously large equivalent 
width and the absence of the other member of the doublet suggest that 
these features have probably been affected 
by the subtraction of telluric Na~I emission.
Two narrow absorption lines (I~10 and I~11 in Figure 1) remain 
unidentified.

\section{SUMMARY AND CONCLUSIONS}

Thanks to gravitational magnification by a factor of $\sim 30$, 
MS~1512$-$cB58 offers a unique insight into the physical properties of 
star forming galaxies at high redshift. We have used LRIS on the Keck 
telescope to secure an intermediate resolution (0.8~\AA), high S/N ratio
($\sim 40$) spectrum covering the wavelength interval 
1145--1930~\AA\ in the rest frame of this galaxy 
($z_{\rm stars} = 2.7268$). 
The observations have revealed a wealth of spectral features from the 
stars and the interstellar medium of cB58, as well as foreground 
galaxies and the intergalactic medium along the line of sight.
From the analysis of these data we arrive at the following main results:

1. Spectral synthesis models based on libraries of O and B stars are 
remarkably successful in reproducing the observed stellar spectrum.
Evidently, the ultraviolet spectral properties of at least 
this high redshift galaxy are very similar to those of local starbursts.
The P-Cygni profiles of C~IV and N~V are best reproduced by a continuous 
star formation model with a Salpeter 
IMF which extends to beyond 50~$M_{\odot}$; we can exclude 
both a flatter IMF and an IMF lacking in the most massive stars.

2. Both stars and gas show evidence for a relatively high degree of metal 
enrichment---we estimate the metallicity to be $\approx 1/4$ of solar. 
This value is $\approx 3$ times higher than the typical metallicity
of damped \lya\ systems at the same redshift (Pettini et al. 1997).
This finding is not surprising, given that we are viewing directly
a region of active star formation, and is consistent with the 
proposal that damped \lya\ systems may preferentially trace  
diffuse gas where star formation proceeds more slowly than in the compact, 
high density regions we see as Lyman break galaxies (Mo, Mao, \& White 
1999; Pettini et al. 1999).

3. The ultraviolet continuum is redder than that of the OB stars whose 
spectral signatures we see directly, probably 
as a result of dust extinction; we deduce 
$E$($B-V$)$ = 0.1 - 0.3$, depending on the 
shape of the extinction curve. 
The implied dust-to-gas ratio is a few times larger than expected, 
suggesting that we have either underestimated the metallicity
or that most of the gas is in ionized and/or molecular form.

4. The relative velocities of interstellar absorption lines, stellar 
photospheric lines, H~II region emission lines, as well as the 
highly asymmetric profile of the \lya\ emission line, 
are all consistent with a 
picture in which the mechanical energy deposited by the starburst
has produced a shell of swept up interstellar matter which is
expanding with a 
velocity of $\sim 200$~km~s$^{-1}$.
We estimate a mass outflow rate, 
$\dot{M}  \approx 60~M_{\odot}~{\rm yr}^{-1}$, 
which is comparable to the star formation rate,
${\rm SFR}_{\rm cB58} \approx 40~M_{\odot}~{\rm yr}^{-1}$,
deduced from the UV luminosity 
($H_0 = 70$~km~s$^{-1}$~Mpc$^{-1}$; $q_0 = 0.1$)
corrected for dust extinction and 
gravitational magnification.
Such galactic winds, which seem to be a common feature of 
star forming galaxies at all redshifts, could 
be the mechanism which regulates star formation,
distributes the metals over large volumes, and 
allows the escape of ionizing photons into the intergalactic medium.

5. Among the intervening absorption we find two
Mg~II systems---one of which is likely to be damped---indicating
the presence of galaxies at $z = 0.829$ and $1.339$ 
close to the sight-line to cB58.

Future observations of cB58 will undoubtedly include 
the familiar rest frame optical emission lines from H~II regions, 
which fall in the H and K near-infrared bands.  
In the rest frame UV, longer exposures should lead to the detection of 
C~III]~$\lambda\lambda 1907,1909$ and O~III]~$\lambda\lambda 1661,1666$
(Garnett et al. 1995). These nebular lines will provide 
measurements of abundances and reddening from the ionized 
gas. Higher resolution spectra, which are feasible with a dedicated 
effort, may allow the determination of the relative abundances of 
different elements and offer chemical clues to the previous 
star formation history. For example, an enhancement of the 
alpha elements relative to Fe-peak elements ([S/Zn] would probably be 
the best probe of this effect), would be an indication of a rapid timescale
for metal enrichment, possibly linking galaxies like cB58 to today's 
bulges. With higher spectral resolution it should also be possible
to search for absorption by molecular hydrogen whose level populations 
are sensitive to temperature, density, and the intensity of the far-UV 
radiation field. Resolving the fine-structure levels of C~I will
give a direct measure of pressure in the neutral component of the ISM.

Finally, the data presented here highlight the 
power of high resolution studies of Lyman break galaxies.
The ability to probe deeply into the physics and chemistry of 
these objects is a
strong motivation for future targeted searches to 
identify other examples of distant galaxies gravitationally 
lensed by foreground clusters.

\acknowledgements

It is a pleasure to acknowledge the many people responsible for
building and maintaining the W.~M.~Keck telescopes 
and the Low Resolution Imaging Spectrograph, and Mindy Kellogg
who kindly helped with the observations.
We thank Claus Leitherer for generously sharing his knowledge of 
starburst galaxies and spectral synthesis models with us.
This work has benefited significantly from numerous conversations with 
colleagues, particularly Ian Howarth, Danny Lennon, Piero Madau,
Bernard Pagel, Blair Savage, Linda Smith, and Pete Storey.
We are grateful to Jim Lawler and Steve Federman for 
communicating results on the $f$-values of Ni~II transitions in advance 
of publication. 
CCS acknowledges support from the U.S. National Science Foundation through
grant AST 94-57446, and from the David and Lucile Packard Foundation.
MG has been supported through grant HF-01071.01-94A from the Space Telescope
Science Institute, which is operated by the Association of Universities for
Research in Astronomy, Inc. under NASA contract NAS 5-26555.


%
%

\begin{deluxetable}{lllll}
\tablewidth{14cm}
\tablecaption{STELLAR PHOTOSPHERIC LINES}
\tablehead{
\colhead{Line} & \colhead{Ion} & \colhead{$\lambda_{\rm lab}^a$ (\AA)}
& \colhead{$\lambda_{\rm obs}^b$ (\AA)} 
& \colhead{$z_{\rm stars}^b$} 
}
\startdata
1  & Si III & 1294.543  & 4825.88 & ~~~~~~2.728 \\
2a & C III  & 1296.33   &         &       \\
2b & Si III & 1296.726  & \raisebox{1.5ex}[0cm][0cm]{{\huge \}} 4832.81} & \raisebox{1.5ex}[0cm][0cm]{~~~~~~2.727} \\
3a & C II   & 1323.929 &         &       \\
3b & N III  & 1324.316  & \raisebox{1.5ex}[0cm][0cm]{{\huge \}} 4934.71} & \raisebox{1.5ex}[0cm][0cm]{~~~~~~2.727} \\
4  & O IV   & 1343.354  & 5005.59 & ~~~~~~2.726 \\
5  & Si III & 1417.237  & 5282.53 & ~~~~~~2.727 \\
6  & C III  & 1427.85   & 5321.24 & ~~~~~~2.727 \\
7  & S V    & 1501.76   & 5594.59 & ~~~~~~2.725 \\
8  & N IV   & 1718.551  & 6403.82: & ~~~~~~2.726: \\
\\
Mean &   &           &         & $2.7268 \pm 0.0008$ \\
\enddata
\tablenotetext{a}{Vacuum wavelengths}
\tablenotetext{b}{Vacuum heliocentric}
\end{deluxetable}

%
%

\begin{deluxetable}{lllllll}
\tablewidth{17cm}
\scriptsize
\tablecaption{INTERSTELLAR ABSORPTION LINES}
\tablehead{
\colhead{Line} & \colhead{Ion} & \colhead{$\lambda_{\rm lab}^a$ (\AA)}
& \colhead{$\lambda_{\rm obs}^b$ (\AA)} 
& \colhead{$z_{\rm abs}^b$}
& \colhead{$W_0^c$ (\AA)} 
& \colhead {Comments}
}
\startdata
1    & Si II & 1190.4158  & 4433.07: & ~~~~~2.7240:  & \ldots &Blended \\
2    & Si II & 1193.2897  & 4444.10: & ~~~~~2.7242:  & \ldots &Blended \\
3    & N I   & 1199.9674  & 4468.80  & ~~~~~2.7241   & 1.71 A &Multiplet\\
4    & Si III & 1206.500  & 4493.62  & ~~~~~2.7245   & 2.83 A &\\
5    & H I   & 1215.6701  & 4527.16  & ~~~~~2.7240   & \ldots &Blended \\
6    & N V   & 1238.821   & 4613.64  & ~~~~~2.7242   & \ldots &Blended \\
7    & S II  & 1253.811   & 4669.64  & ~~~~~2.7244   & 0.45 A & \\
8    & S II  & 1259.519   &           &          &        & \\
9    & Si II & 1260.4221  & \raisebox{1.5ex}[0cm][0cm]{{\huge \}} 4693.48}  & \raisebox{1.5ex}[0cm][0cm]{~~~~~\ldots}   & \raisebox{1.5ex}[0cm][0cm]{2.87 A}&\\
10   & Si II* & 1264.7377 & 4709.15  & ~~~~~2.7234   & 0.55 A &\\
11   & C I   & 1277.4626  & 4755.64  & ~~~~~2.7227   & 0.33 A &\\
12   & O I   & 1302.1685  & 4849.61: & ~~~~~2.7243:  &        & \\
13   & Si II & 1304.3702  & 4857.66: & ~~~~~2.7241:  & \raisebox{1.5ex}[0cm][0cm]{{\huge \}} 4.39 A} &\\
14   & Ni II & 1317.217   & 4906.66  & ~~~~~2.7250   & 0.20 B & Blended? \\
15   & C II  & 1334.5323  & 4970.22  & ~~~~~2.7243   & 3.46 A &\\
16   & Ni II & 1370.132   & 5102.58  & ~~~~~2.7241   & 0.27 A &\\
17   & Si IV & 1393.755   & 5190.88  & ~~~~~2.7244   & 1.94 A &\\
18   & Si IV & 1402.770   & 5224.01  & ~~~~~2.7241   & 1.42 A &\\
19   & Si II & 1526.7066  & 5685.56  & ~~~~~2.7241   & 2.72 A &\\
20   & Si II* & 1533.4312  & 5710.34  & ~~~~~2.7239   & 0.59 A &\\
21   & C IV  & 1548.195   & 5765.52  & ~~~~~2.7240   &        &\\
22   & C IV  & 1550.770   & 5775.88  & ~~~~~2.7245   & \raisebox{1.5ex}[0cm][0cm]{{\huge \}} 3.8 B }& \raisebox{1.5ex}[0cm][0cm]{Blended with stellar C~IV }\\
23   & Fe II & 1608.4511  & 5990.21  & ~~~~~2.7242   & 1.12 A & \\
24   & Al II & 1670.7874  & 6221.37  & ~~~~~2.7236   & 2.81 A & \\
25   & Ni II & 1709.600   & 6366.02  & ~~~~~2.7237   & 0.29 C & \\
26   & Ni II & 1741.549   & 6485.67  & ~~~~~2.7241   & 0.42 C & \\
27   & Ni II & 1751.910   & 6527.40  & ~~~~~2.7259   & 0.24 C & \\
28   & Si II & 1808.0126  & 6733.75  & ~~~~~2.7244   & 0.51 B & \\
29   & Al III & 1854.7164 & 6907.87  & ~~~~~2.7245   & 1.72 B & \\
30   & Al III & 1862.7895 & 6938.09  & ~~~~~2.7246   & 1.38 B & \\
\\
Mean &        &           &           & $2.7242 \pm 0.0005$   \\
\enddata
\tablenotetext{a}{Vacuum wavelengths}
\tablenotetext{b}{Vacuum heliocentric}
\tablenotetext{c}{Rest frame equivalent width and $1 \sigma$ error. 
A: Error $\leq 10$\%; 
B: Error $\leq 20$\%;
C: Error $\leq 30$\%.} 
\end{deluxetable}

%
%

\hoffset -0.5cm
\begin{deluxetable}{lllllllll}
\tablewidth{17cm}
\footnotesize
\tablecaption{INTERSTELLAR ABUNDANCES}
\tablehead{
\colhead{Line} & \colhead{Ion} & \colhead{$\lambda_{\rm lab}^a$ (\AA)}
& \colhead{$f$}
& \colhead{Ref.$^b$}
& \colhead{log $N$ (cm$^{-2}$)}
& \colhead{log (X/H)}
& \colhead{log (X/H)$_{\odot}^c$}
& \colhead{[X/H]$_{\rm cB58}^d$}
}
\startdata
5    & H I   & 1215.6701  & 0.4164  & 1 & 20.88 & \ldots  & \dots   & \ldots \\
\\
28   & Si II & 1808.0126  & 0.00218 & 2 & 15.91 & $-4.97$ & $-4.45$ & $-0.52$ \\
\\
7    & S II  & 1253.811   & 0.01088 & 1 & 15.47 & $-5.40$ & $-4.73$ & $-0.68$\\
\\
14   & Ni II & 1317.217   & 0.0774: & 1 & 14.23 &         &         &        \\
16   & Ni II & 1370.132   & 0.0765  & 3 & 14.33 &         &         &        \\
25   & Ni II & 1709.600   & 0.0348  & 4 & 14.51 &         &         &        \\
26   & Ni II & 1741.549   & 0.0419  & 4 & 14.57 &         &         &        \\
27   & Ni II & 1751.910   & 0.0264  & 4 & 14.53 &         &         &        \\
     & Ni II & & & & $14.45^{+0.12}_{-0.17}$    & $-6.43$ & $-5.75$ & $-0.68$\\
\enddata
\tablenotetext{a}{Vacuum wavelengths}
\tablenotetext{b}
{References for $f$-values --- 1: Morton (1991); 2: Bergeson \& Lawler (1993); 
3: Zsarg\'{o} \& Federman (1998); 4: Fedchak \& Lawler (1999).}
\tablenotetext{c}{Solar (meteoritic) abundances from the compilation 
by Anders \& Grevesse (1989).}
\tablenotetext{d}{[X/H]$_{\rm cB58}$ = log (X/H) $-$ log (X/H)$_{\odot}$.}\\
\end{deluxetable}

%
%

\begin{deluxetable}{lllllll}
\tablewidth{17cm}
\tablecaption{WEAK EMISSION LINES}
\tablehead{
\colhead{Line} & \colhead{Ion} & \colhead{$\lambda_{\rm lab}^a$ (\AA)}
& \colhead{$\lambda_{\rm obs}^b$ (\AA)} 
& \colhead{$z_{\rm em}^b$}
& \colhead{$W_0^c$ (\AA)} 
& \colhead {Comments}
}
\startdata
1    & Si II* & 1264.7377 & 4716.45  & ~~~~~2.729   & $0.165 \pm 0.015$  &\\
2    & Si II* & 1309.2757 & 4881.28  & ~~~~~2.728   & $0.070 \pm 0.015$  &\\
3    & C II*  & 1335.7077 & 4983.12  & ~~~~~2.731   & $0.110 \pm 0.015$  &\\
4    & N IV]  & 1486.496  & 5545.20: & ~~~~~2.730:  & $0.085 \pm 0.020$  & Noisy \\
\\
Mean &        &           &           & $2.7295 \pm 0.001$   \\
\enddata
\tablenotetext{a}{Vacuum wavelengths}
\tablenotetext{b}{Vacuum heliocentric}
\tablenotetext{c}{Rest frame equivalent width and $1 \sigma$ error from 
counting statistics only (no continuum uncertainty included)} 
\end{deluxetable}

%
%
\begin{deluxetable}{lccc}
\tablewidth{0pc}
\tablecaption{RELATIVE VELOCITIES IN cB58}
\tablehead{
\colhead{Spectral Features} & \colhead {No. of Lines} & \colhead{$z^a$} &
\colhead{$v$ (km~s$^{-1}$)$^b$} 
}
\startdata
Stellar photospheric lines & 8  & $2.7268 \pm 0.0008$ & 0.00 \\
Interstellar abs. lines    & 29 & $2.7242 \pm 0.0005$ & $-210 \pm 80$ \\
H~II emission lines        & 4  & $2.7296 \pm 0.0012$ & $+230 \pm 110$\\
Ly$\alpha$ emission line   & 1  & $2.7326$            & $+465$ \\
\enddata
\tablenotetext{a}{Vacuum heliocentric}
\tablenotetext{b}{Relative to $z_{\rm stars}$}
\end{deluxetable}

%
%

\begin{deluxetable}{llllll}
\tablewidth{14cm}
\footnotesize
\tablecaption{INTERVENING ABSORPTION LINES}
\tablehead{
\colhead{Line} & \colhead{$\lambda_{\rm obs}^a$ (\AA)} & 
\colhead{Identification} & \colhead{$z_{\rm abs}^a$} & 
\colhead{$W_0^b$ (\AA)} & \colhead{Comments} 
}
\startdata
1  & 5113.79  & Mg II 2796.352   & ~~~~~0.8287   & 1.19 A &\\
2  & 5127.06  & Mg II 2803.531   & ~~~~~0.8288   & 0.96 A &\\
\hline
\\
3  & 5483.90  & Fe II 2344.214   & ~~~~~1.3393   & 0.33 A &\\
4  & 5573.37  & Fe II 2382.765   & ~~~~~1.3390   & 0.62 A &\\   
5  & 6050.20  & Fe II 2586.6500  & ~~~~~1.3390   & 0.59 B & Blended\\
6  & 6082.13  & Fe II 2600.1729  & ~~~~~1.3391   & 0.45 B &\\
7  & 6541.22  & Mg II 2796.352   & ~~~~~1.3392   & 0.96 A &\\
8  & 6558.47  & Mg II 2803.531   & ~~~~~1.3394:  & 0.61 B &\\
9  & 6672.79  & Mg I  2852.9642  & ~~~~~1.3389   & 0.42 C & \\
\\
Mean &        &                  & $1.3391 \pm 0.0002$ & & \\
\hline
\\
10 & 5158.21  & ?                & ~~~~~?        & 1.24$^c$ A  & Unidentified\\
11 & 5666.56  & ?                & ~~~~~?        & 0.99$^c$ A  & Unidentified \\
12 & 5891.54  & Na I 5891.5833   & ~~~~~0.0000   & 0.97:       & Galactic ISM? \\
   &          &                  &               & &(affected by sky emission)
\enddata
\tablenotetext{a}{Vacuum heliocentric}
\tablenotetext{b}{Rest frame equivalent width and $1 \sigma$ error. 
A: Error $\leq 10\%$;
B: Error $\leq 20\%$;
C: Error $\leq 30\%$.}
\tablenotetext{c}{{\it Observed} frame equivalent width}
\end{deluxetable}

\newpage

%
%

\begin{figure}
\figurenum{1}
\vspace*{-2cm}
\hspace*{+0.25cm}
\psfig{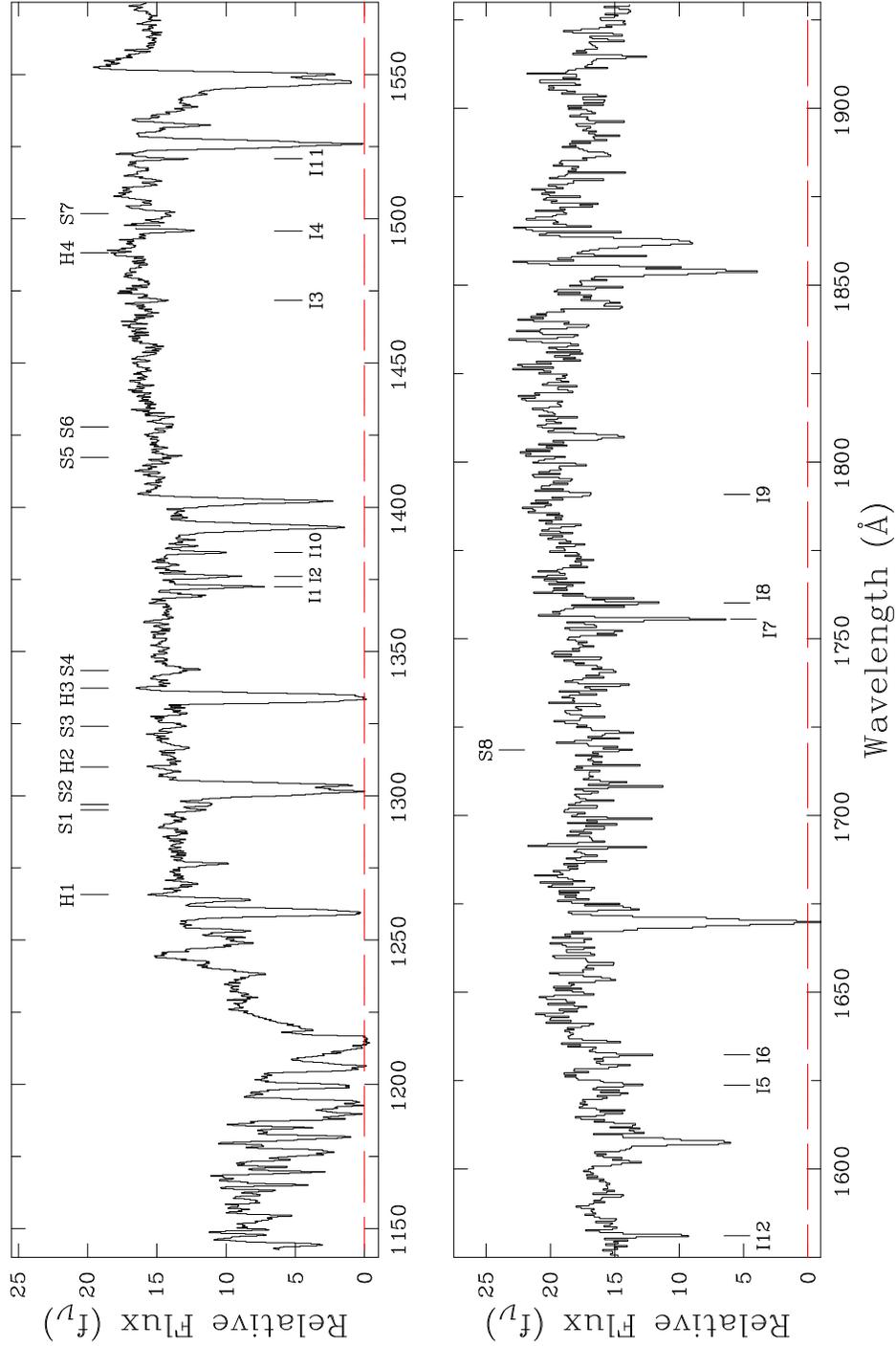}
\vspace{-0.5cm}
\figcaption{LRIS spectrum of MS~1512$-$cB58 reduced to the
systemic redshift of the galaxy, $z_{\rm star} = 2.7268$. Tick
marks above the spectrum identify weak
stellar lines (labelled S$n$) listed in Table 1,
and weak emission lines (labelled H$n$---see Table 4)
which we attribute to H~II gas.
Tick marks below the spectrum
(labelled I$n$) mark the positions of the intervening absorption
lines listed in Table 6. The numerous interstellar lines in
MS~1512$-$cB58 have not been marked to
avoid crowding in the figure, but can be readily recognized by
reference to Table 2.}
\end{figure}

%
%

\begin{figure}
\figurenum{2}
\vspace*{-2.5cm}
\hspace*{-1.5cm}
\psfig{figure=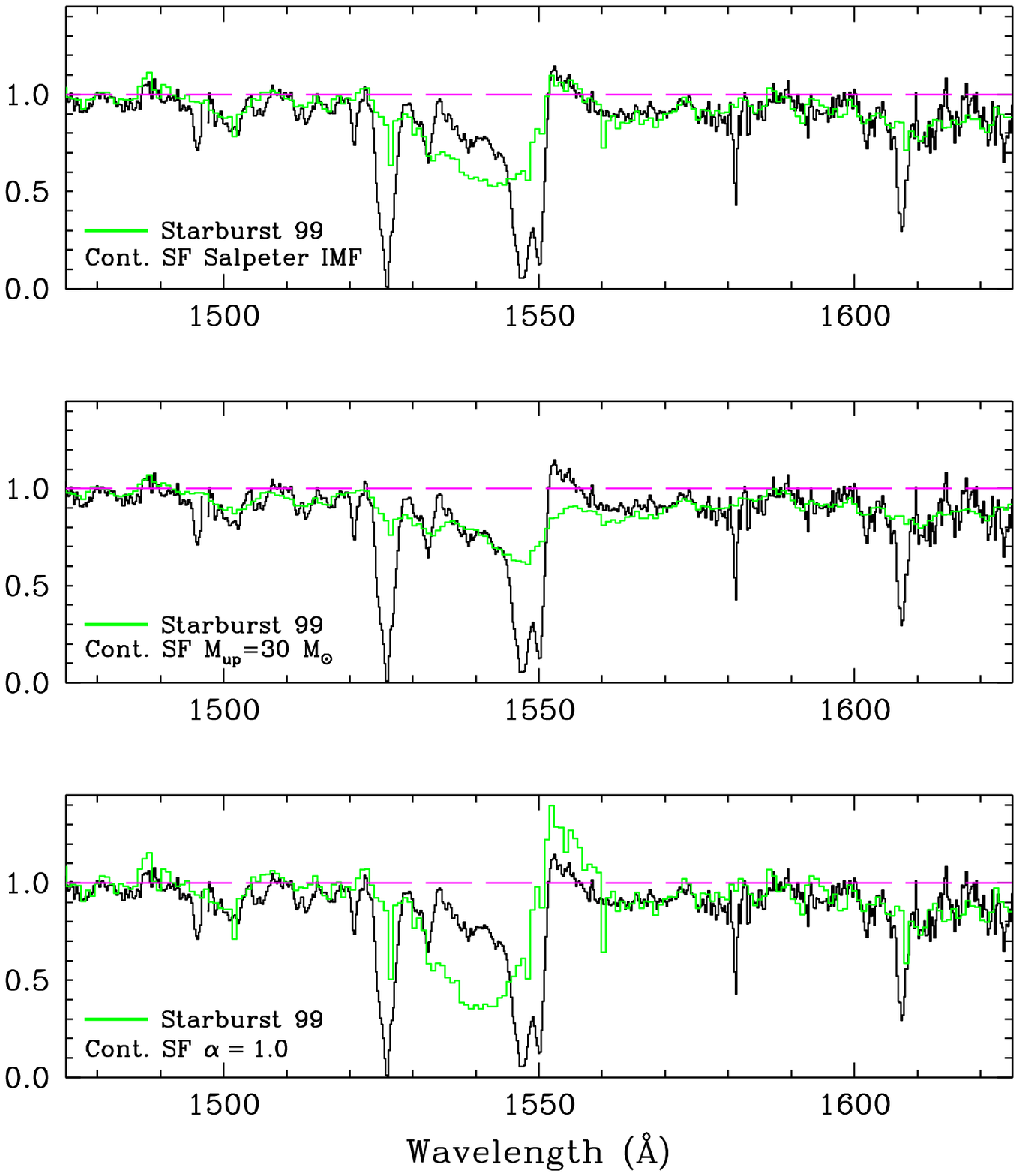,width=190mm}
\vspace{-4.5cm}
\figcaption{Comparison between the observed spectrum of cB58 
(black histogram) in the region of the C~IV~$\lambda 1549$ line
with different spectral synthesis models (coloured 
histograms, or light grey in the black and white version of the paper), 
as indicated. See text for discussion. The $y$-axis is residual intensity.}
\end{figure}

%
%

\begin{figure}
\figurenum{3}
\vspace*{-2.5cm}
\hspace*{-1.5cm}
\psfig{figure=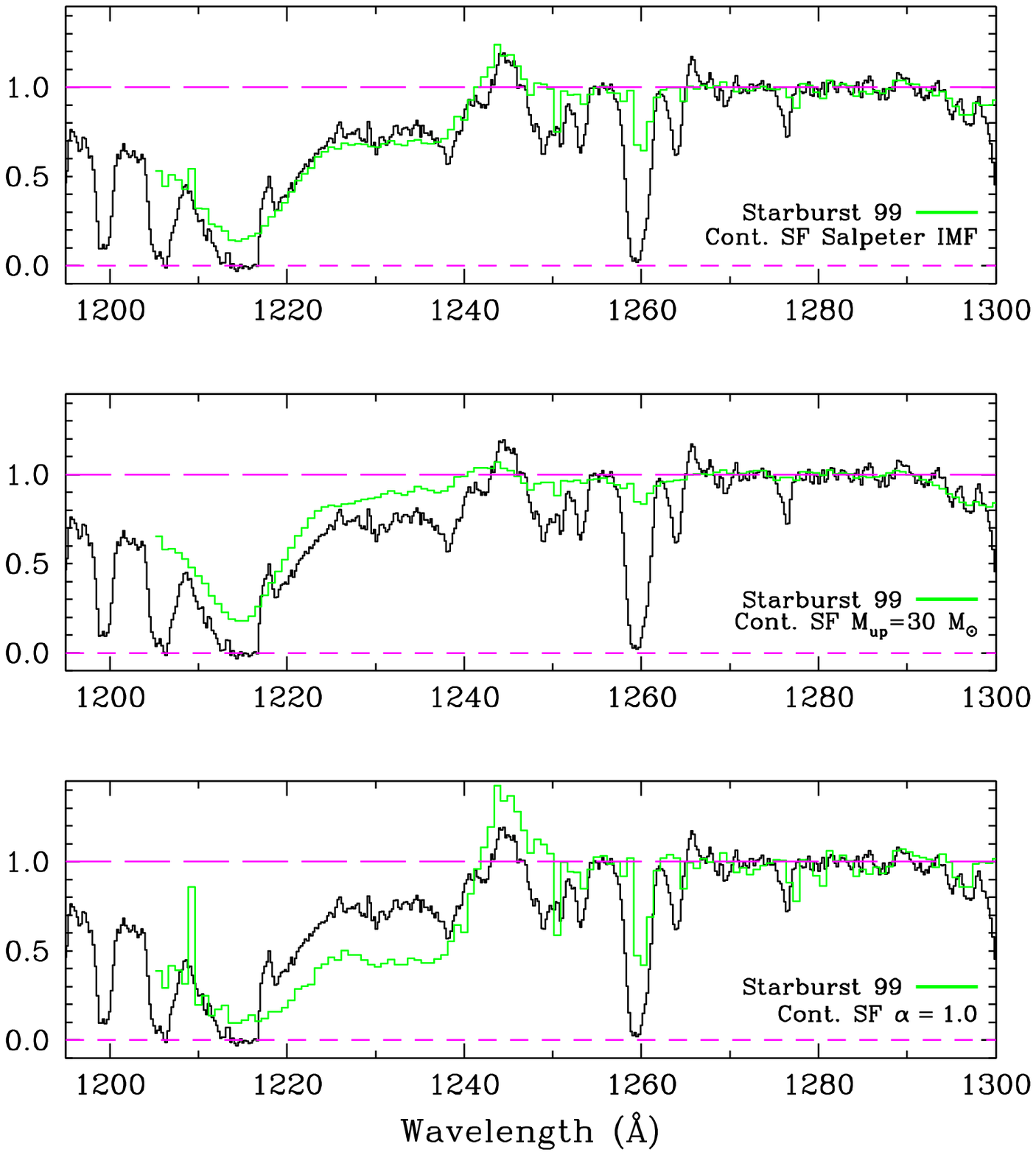,width=190mm}
\vspace{-4.5cm}
\figcaption{Comparison between the observed spectrum of cB58 
(black histogram) in the region of the N~V~$\lambda 1240$ line
with different spectral synthesis models (coloured 
histograms, or light grey in the black and white version of the paper), 
as indicated. See text for discussion. The $y$-axis is residual intensity.}
\end{figure}

%
%

\begin{figure}
\figurenum{4}
\vspace*{-3.5cm}
\hspace*{-1.5cm}
\psfig{figure=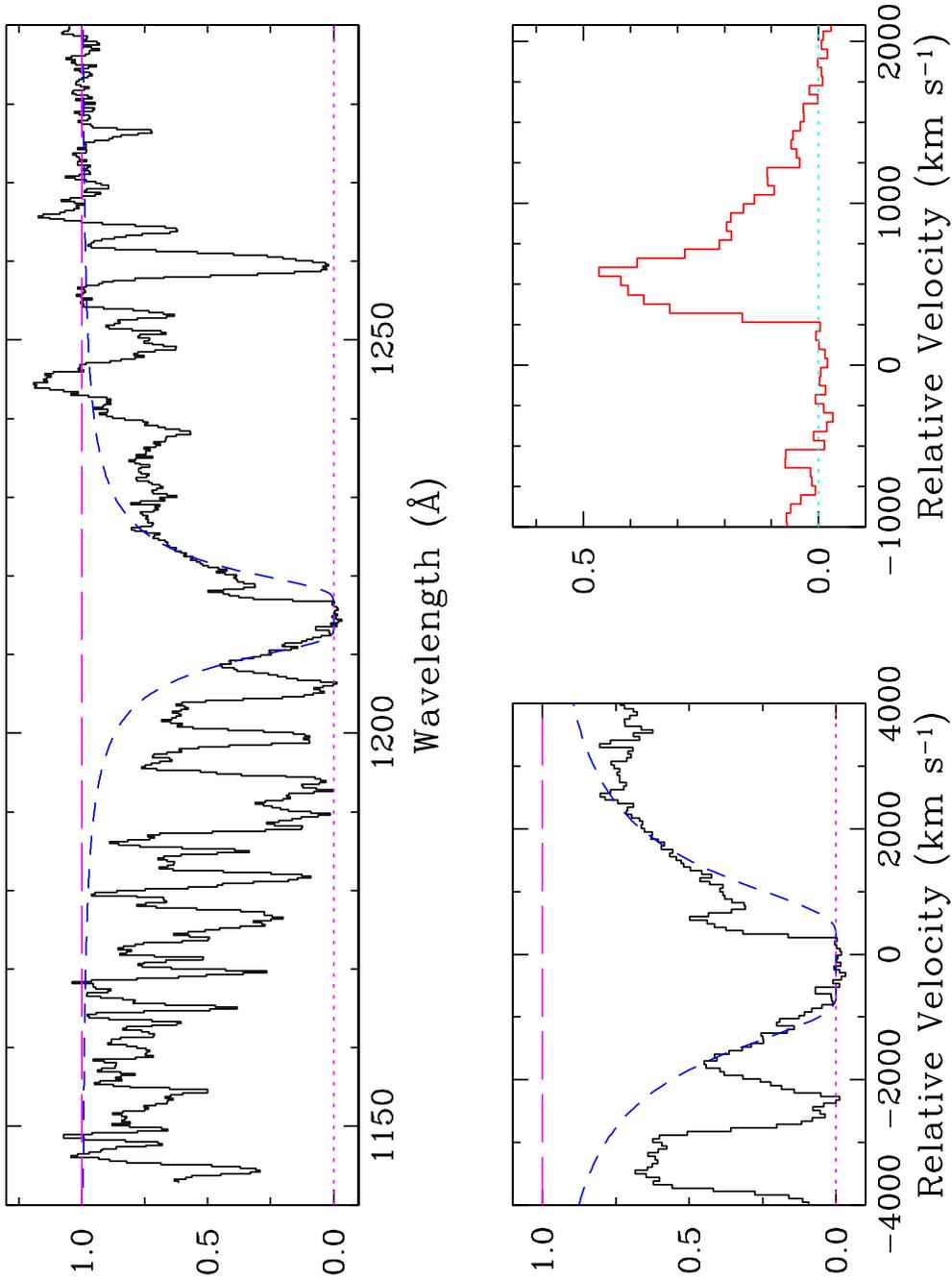,width=200mm}
\vspace{-5.0cm}
\figcaption{Decomposition of the \lya\ profile in cB58. {\it Top 
panel}---Black histogram: observed spectrum; short dash line: theoretical 
absorption profile for $N$(H~I)$ = 7.5 \times 10^{20}$~cm$^{-2}$. 
{\it Bottom left-hand panel}---Same as the top panel, but on a 
velocity scale relative to $z_{\rm stars}$. 
{\it Bottom right-hand panel}---Residual \lya\ emission when the
absorption component is subtracted out. In each case the $y$-axis is residual 
intensity.}
\end{figure}

\end{document}